# ON THE PROBLEM OF EXTENDED SPECIAL RELATIVITY CREATION [*]

V. V. Korukhov [a], O. V. Sharypov [b]

[a] Institute of Philosophy and Law, Siberian Branch of Russian Academy of Sciences
(Nikolayev st., 8, Novosibirsk, 630090, Russia. kvv@philosophy.nsc.ru)

[b] Institute of Thermophysics, Siberian Branch of Russian Academy of Sciences
(Lavrentyev av., 1, Novosibirsk, 630090, Russia. sharypov@itp.nsc.ru)

This paper presents an approach to the creation of a variant of Extended Special Relativity that takes into consideration the existence of limiting relativistically invariant quantities (Planck parameters). It shows the possibility of excluding unphysical predictions of relativity theories thanks to the use of the concept of the maximum velocity of the observed motion of objects. It proposes a model of a vacuum-like medium with a kinematical property of relativistically invariant rest. The Planck quantities are considered as fundamental physical constants related to the structure of this medium.

## 1. Introduction

At present, theoretical results got within the framework of different approaches have proved the limit nature of some Planck quantities, such as speed, action, length, and time [1-4], mass and density [5, 6], electric field potential, acceleration, and temperature [7-15], electric resistance and entropy [16], etc. These results have been generalized, which has led to a new methodological principle of developing fundamental post-non-classical theories called "$\hbar c G$-principle" [16, 17]. According to this principle, the Planck values of physical quantities must play the part of relativistically invariant limits (in appropriate processes and phenomena). In other words, for all physical quantities their ranges of values are limited by finite (not equal to infinity or zero) universal limits, and the values of these limits do not depend on the chosen inertial reference frame (IRF). Relativistic invariance of the Planck quantities $(X_{\text{pl}})$ follows directly from their definitions, as each of them is expressed in terms of a combination of certain fundamental physical constants (FPC) – $\hbar$, $c$, $G$, $k$ – in compliance with the dimension: $X_{\text{pl}} \equiv \hbar^{\chi} \cdot c^{\eta} \cdot G^{\theta} \cdot k^{\sigma}$. According to the definition all Planck quantities are considered to be FPC, equally with the Planck constant $(\hbar)$, the speed of light in empty space $(c)$, gravitation constant $(G)$, and the Boltzmann constant $(k)$.

In [16] it was first drawn attention to the property of the relativistic invariance (observer independence) of the Planck quantities. Owing to this property, the Planck values of physical quantities become absolute, which differs them qualitatively from relative values characterizing the states of material objects [18].

Taking into consideration relativistic invariance and limit nature of the Planck quantities plays a crucial role in development of physical theories. It is illustrated by the changes accompanying transitions from classical mechanics to:
- Special Relativity (SR) on account of the Planck speed $c$;
- Non-relativistic Quantum Mechanics on account of the Planck constant $\hbar$;
- Relativistic Quantum Mechanics on account of two Planck quantities $\hbar$, $c$.

---

[*] The work was supported by Russian Foundation for Basic Researches, project No 03-06-80367.

The same is characteristic for the transition from Newtonian classical Gravitation theory to the relativistic one – General Relativity – on account of the finiteness of the speed of light *c*. We should note that a theory that would take into consideration the limit nature of the gravitation constant *G* (as the Planck quantity) has not been developed yet[1].

The choice of a certain set of the Planck quantities determines the qualitative specification (the content) of the theory. Taking into consideration more Planck quantities causes the necessity to develop an extended theory, which generalizes the preceding one in accordance with the correspondence principle. Therefore, it's possible to develop new fundamental theories, extending the bases of the existing ones by inclusion of a certain set of the Planck quantities. This way of gradual generalization leads to the creation of a Unified theory which will include all Planck quantities (all FPS) and serve as an ideal of the post-non-classical unity of physics.

Attempts to extend SR is one of particular trends of this process. In the last few years there have been published a lot of works whose authors consider variants of SR modification introducing one or another observer independent quantity into SR: the Planck length (momentum, energy) [22-25 and others]. All these approaches are joined by the common name of "Doubly Special Relativity" which emphasizes the fact that unlike Einstein's SR the new constructions use not only one, but two Planck quantities. The paper [26] presents an attempt to develop a theory considering three invariant quantities ("Triply Special Relativity").

Apparently, generalization of SR must result in a theory that will comprise the limit and invariant values of all quantities used in mechanics: length, time, mass, momentum, acceleration, force, energy, etc. It is suggested in [27] to call such a theory "Extended Special Relativity" (ESR). The development of this theory should eliminate unphysical predictions of SR leading to unrestrictedly large and small (i.e. infinite and zero) values of physical quantities. Thus the field of applicability of Einstein's SR will be bounded, – the new theory must predict the finite limits for the range of values of physical quantities, and we believe that these limits will be determined by the Planck parameters.

## 2. Discrete-continuous structure of space-time

Since the 1930s physicists discuss the concept of the fundamental length. A number of characteristic extents arising in micro world theories have been "cast" for the part of the fundamental length $(L)$. According to modern theoretical views, the Planck length $l_{pl} = \sqrt{\hbar G/c^3} \sim 10^{-33}$ cm can play the role of fundamental length. Thereby it is assumed that the length $L = l_{pl}$ possesses the property of observer independence and limitness to all material objects and processes (universality), regardless of their nature [20, 28].[2]

Obviously, a theory based on the concepts of the fundamental length $l_{pl}$ and invariant velocity *c* should also include the concept of the fundamental duration: $t_{pl} = l_{pl}/c \sim 10^{-44}$ sec. The values $l_{pl}$, $t_{pl}$ can be used to develop one or another variant of ESR. Since SR can be considered as the "theory of space-time", the introduction of the concepts of the minimum invariant length and time into it means a change in the notion of the structure of space-time in the small scales. The quantities $l_{pl}$ and $t_{pl}$, that are not equal to zero, attribute the property of discreteness (discontinuity) to the structure of space and time. On the other hand, the relativistic invariance of the values $l_{pl}$ and $t_{pl}$ reproduces the absolute properties of the zero element of a continuous set[3]. It can be concluded that the structure of such space (and time) is neither discrete nor continuous in the strict sense. There are no preferential directions (isotropy) in a space with this kind of structure, all IRF are physically equivalent. The term "discrete-continuous structure" was introduced to define the specific character of the structure of space (time) with $l_{pl}$ $(t_{pl})$ [17, 20].

---

[1] [17, 19] present a hypothesis according to which the classical Gravitation theory can be generalized on the basis of the postulate of the gravitation constant being limiting and invariant quantity: $G = \left(\rho_{pl} \cdot t_{pl}^2\right)^{-1} \leq \left(\Delta\rho \cdot \Delta t^2\right)^{-1}$ (similar to the Heisenberg's uncertainty relations in Quantum Mechanics [20, 21]), where ρ is density, *t* is time.

[2] In this sense, the fundamental length is an absolute unit. Relativistic invariance of the Planck length was first emphasized in [16].

Apparently, a space with a discrete-continuous structure possesses specific topological properties. Metric properties of a discrete-continuous space in the small scales must be different from those of the flat space with continuous structure (Euclidean space), and in the large scales, in comparison with $l_{pl}$ $(t_{pl})$, they should approximate to them. This kind of space turns out to be curved in the small scales, unlike Riemannian spaces.

The present paper will use a simplified model of discrete-continuous space-time. As the first approximation we will assume that in the scales larger than $l_{pl}$ and $t_{pl}$ space-time is flat and possesses the continuous structure. The quantities $l_{pl}$ and $t_{pl}$ bound the physically permissible values of length and time. This model corresponds to the introduction of observer independent cut-off parameters to the model of space-time used in SR.

### 3. The concept of the maximum velocity of moving objects

The new conception of space and time structure dealing with the presence of limitary small, invariant length $(l_{pl})$ and time $(t_{pl})$ necessitates revision of the concept of mechanical velocity of objects in motion.

In space and time with a continuous structure (i.e. when $l_{pl} = 0$ and $t_{pl} = 0$) the value of velocity of motion is defined in every point of the trajectory as $V = dr/dt$, where $dr$ and $dt$ are infinitesimal changes of coordinate and time. Introduction of certain limitary small scales $l_{pl}$ and $t_{pl}$ will obviously make it necessary to adjust the concept of velocity, and the amendment must be connected with the quantity $l_{pl}$.

Hereinafter we bound the range of phenomena under study to the field of elementary particles. It will allow us to change from coordinate space to momentum one whenever necessary. Therewith we will conceive motion in the space-time corresponding to the macroscopic experience. The inclusion of the Planck parameters will result in a definite limitation of the range of the values of physical quantities.

Extending SR due to inclusion of the fundamental length $(l_{pl} \neq 0)$, which is defined with the help of the Planck constant $\hbar$, of course does not mean automatic transition to quantum notions. However, the new theory should take into consideration the limit character of $l_{pl}$ (and other FPC), and this property of the fundamental length is directly connected with the limit nature of the quantum of action. The given property of $\hbar$ is expressed by the Heisenberg's uncertainty relations that serve as the basis of quantum mechanics: $\Delta p \cdot \Delta l \geq \hbar$ ($\Delta p$ is a change of momentum). If ESR considers $l_{pl}$ and pretends to describing kinematics up to microscopic (Planck) space-time scales it should accord with the Heisenberg's uncertainty relations.

Suppose that an object (an elementary particle with the rest mass $m_0$) is at rest in the inertial frame that is moving at a constant speed $V$ in relation to an observer. Then the minimum range of localization of the particle in the observer's system $l$ is connected with the uncertainty of the value of its momentum (or velocity $\Delta p/m_0$). Let $V \to V_{max} \neq \infty$, where $V_{max}$ is a certain

---

[3] Such quantities act as a "finite", or to be more accurate, "*actual*" zero of a set [18].

*limit* value of the velocity (introduction of which we want to ground) of the observed motion of a concrete object. Then, according to non-relativistic quantum conceptions, $p \to p_{max} = m_0 V_{max}$ and $l \sim \hbar/(m_0 V_{max}) \neq 0$ $(\Delta p \sim p)$. From the point of view of relativistic quantum ideas, the minimum range of localization of the particle is not equal to zero and is characterized by the Compton size: $l \sim l_0 = \hbar/(m_0 c) \neq 0$. Contrary to this, in case when $V \to V_{max} = c$ SR gives the result: $l \to 0$. Of course the velocity equal to $c$ cannot characterize motion of a material object in SR and the value $l$ for any object must be always finite. Nevertheless, in accordance with SR, when $V \to c$ the size $l$ becomes infinitesimal, which contradicts the concept of the fundamental length and does not meet the requirements of quantum mechanics. Elimination of these contradictions can be related with consideration of the minimum elements of space-time structure (the limit nature of the Planck quantities) in ESR. Respectively, the concept of the allowable speed of substance objects should also change.

It can be noted that SR contains a statement, which says that the maximum velocity of motion of material objects $(V_{max})$ in any case must be less then $c$. But SR does not determine a finite difference between $V_{max}$ and $c$: $(c^2 - V_{max}^2) \to 0$. The idea of the discrete-continuous structure of space-time gives a reason to *suppose* that there exists a finite limit of this difference $(c^2 - V_{max}^2) > 0$. Let us represent it in the following form:

$$\frac{V^2}{c^2} \leq \frac{V_{max}^2}{c^2} = 1 - \alpha, \tag{1}$$

where $V_{max}$ is the maximum velocity of motion of the object, $V_{max}^2 = c^2(1-\alpha)$, $\alpha \geq 0$ is an indefinite function connected with the fundamental length (and maybe other FPC). The quantity $\alpha$ is a relativistically invariant quantity that can take into consideration the physical parameters of a certain object (for instance, $m_0$) and that according to the correspondence principle must approach to zero when $l_{pl} \to 0$.

One can get the form of the expression (1) on the basis of the following reasoning. The Compton size of the particle $(l_0)$ defines the characteristic time $t_0 = l_0/c$. Let in time $t_0$ the change of coordinate of a given object is $l_1 = Vt_0 \leq V_{max} t_0 = (l_1)_{max}$. In the same period of time the light travels the distance $l_2 = ct_0$. Since $l_2 > (l_1)_{max}$ and $l_{pl}$ has a limit nature, we get:

$$l_2^2 - (l_1)_{max}^2 = \omega^2 l_{pl}^2 \geq l_{pl}^2,$$

where $\omega$ is a constant, $\omega^2 \geq 1$. Then:

$$\frac{V_{max}^2}{c^2} = 1 - \omega^2 \frac{l_{pl}^2}{l_0^2} \equiv 1 - \alpha.$$

Thereby we obtain for the elementary particles:

$$\frac{V^2}{c^2} \leq \frac{V_{max}^2}{c^2} = 1 - \omega^2 \frac{m_0^2}{m_{pl}^2},$$

which coincides in form with the supposition (1). The expression of α will be found below on the basis of limiting relationships of the theory under development.

The idea of the existence of a limit observer independent velocity of motion of a micro object is a new approach to the solution of the problem of physical concordance between the results of SR and the quantum mechanical Heisenberg's uncertainty relations. It also ensures the absence of unphysical predictions of theory, i.e. unrestrictedly large and small volumes of physical quantities [16, 17]. This idea and the definition (1) are possible consequences of a hypothesis of the fundamental length $l_{pl}$ and other limit invariant Planck quantities.

### 4. The consequences of introduction of the concept of the maximum velocity of motion of an object in ESR

The variant of extension of SR under consideration retains Einstein's postulates. The extension of the theory is achieved by introduction of an additional postulate that must not contradict the principles of SR. This methodological requirement to the development of a new theory agrees with the correspondence principle. Thus we base on the relativity principle and the constancy of the speed of light $c$ in all IRF ($c$ is considered to be independent of photon energy). The additional postulate is the statement that there exist observer independent limits – the Planck quantities – that must be taken into account in the new theory. Among them – the fundamental length $l_{pl}$.

Thanks to the choice of postulates, a number of basic relations and statements of SR are retained in the given variant of ESR. First of all, we preserve the conclusion about the invariance of the interval $(s)$ between events in the rest $K$ and moving $K'$ frames:

$$(s_{21})^2 = c^2(t_2 - t_1)^2 - (x_2 - x_1)^2 - (y_2 - y_1)^2 - (z_2 - z_1)^2 = \\ = (s'_{21})^2 = c^2(t'_2 - t'_1)^2 - (x'_2 - x'_1)^2 - (y'_2 - y'_1)^2 - (z'_2 - z'_1)^2, \quad (2)$$

here we assume the limitations imposed by the adopted model of space-time:

$$l_{pl}^2 < c^2(t_2 - t_1)^2;\ (x_2 - x_1)^2;\ (y_2 - y_1)^2;\ (z_2 - z_1)^2;\ (s_{21})^2.$$

The system $K'$ moves in relation to the system $K$ at a constant rate $V$ along the axis $x$. The system $K'$ is connected with a concrete micro object (for instance, an elementary particle with the rest mass $m_0$). Therefore, the range of values of the velocity of the relative motion $V$ is limited by the quantity $V_{max}^2 = c^2(1-\alpha)$, in accordance with definition (1).

Linear transformations of coordinate and time that guarantee invariance of the interval (2) coincide in form with the Lorentz transformations:

$$x = \frac{x' + Vt'}{\sqrt{1 - V^2/c^2}}, \qquad y = y', \qquad z = z', \qquad t = \frac{t' + Vx'/c^2}{\sqrt{1 - V^2/c^2}}. \quad (3)$$

Using (3), we can get the formulae describing the relation between the proper time $t_0 = (t'_2 - t'_1)$ and the time $t = (t_2 - t_1)$ counted by a stationary clock in frame $K$, as well as the relation between the proper length of the object $l_0 = (x_2 - x_1)$ and the length $l = (x'_2 - x'_1)$ measured in the moving frame $K'$:

$$t = \frac{t_0}{\sqrt{1-V^2/c^2}} \quad (\text{as } x'_2 = x'_1,\ y'_2 = y'_1,\ z'_2 = z'_1); \tag{4}$$

$$l = l_0\sqrt{1-V^2/c^2} \quad (\text{as } t'_2 = t'_1). \tag{5}$$

Expressions (4) and (5) are analogous in form to relations in SR. Yet it can be noted that according to (5), when the velocity of relative motion of frames $V$ approaches to the limiting value $V_{max} < c$, the quantity $l$ converges to a *finite* limit:

$$l^2\big|_{V^2 \to V_{max}^2} \to l_0^2\left(1 - \frac{V_{max}^2}{c^2}\right) = \alpha l_0^2. \tag{6}$$

It means that the Lorentz-Fitzgerald contraction in ESR is bounded by a minimum limiting value, unlike SR that predicts zero as the limit. In this way, one of the physically meaningless results is excluded in principle from the theory.

As appears from the postulate about existence of the observer independent limiting small length, when $l_0 \to l_{pl}$ the value of $l$ must also converge to $l_{pl}$. According to (5) and (6) it means that

$$\alpha\big|_{l_0 \to l_{pl}} \to 1 \quad \Leftrightarrow \quad V\big|_{l_0 \to l_{pl}} \to 0. \tag{7}$$

It follows from (4) that for any value of the velocity $V \leq V_{max}$ to the finite proper time $t_0$ there should correspond a finite time $t$ in the rest frame. If $t_0 \to t_{pl}$, then due to the invariance of the quantity $t_{pl}$ the value $t$ must also converge to $t_{pl}$. It is ensured by the condition $\alpha\big|_{t_0 \to t_{pl}} \to 1 \Leftrightarrow V\big|_{t_0 \to t_{pl}} \to 0$ that concords with (7) if $l_0$ and $t_0$ are mutually coordinated quantities, for example the Compton size of the particle and corresponding time $t_0 = l_0/c$.

In SR, basing on expression (2) for the interval, we can define the Lagrange function (**L**) for a free particle, applying the principle of least action $(S)$:

$$S = \int_{t_1}^{t_2} \mathbf{L}\, dt, \qquad \mathbf{L} = -m_0 c^2 \sqrt{1 - \frac{V^2}{c^2}}. \tag{8}$$

The momentum and energy of the particle are defined on basis of (8):

$$p^2 = \frac{m_0^2 V^2}{(1-V^2/c^2)}, \qquad E^2 = \frac{m_0^2 c^4}{(1-V^2/c^2)}. \tag{9}$$

In accordance with ESR postulates, the maximum value of the particle energy is limited by the invariant Planck quantity $E \leq E_{pl}$, where $E_{pl} = m_{pl} c^2 = \sqrt{\hbar c^5/G}$.

As $E_{max}^2 = \dfrac{m_0^2 c^4}{1-V_{max}^2/c^2}$, then:

$$m_{pl}^2 c^4 = \frac{m_0^2 c^4}{1 - V_{max}^2/c^2}.$$

Therefore, for the maximum velocity of the particle we get:

$$V_{max}^2 = c^2 \left(1 - \frac{m_0^2}{m_{pl}^2}\right) = c^2 \left(1 - \frac{l_{pl}^2}{l_0^2}\right).$$

Correspondingly:

$$\alpha = \frac{m_0^2}{m_{pl}^2} = \frac{l_{pl}^2}{l_0^2}, \quad \omega^2 = 1. \tag{10}$$

The same result can be achieved if we demand the length of the object in the observer's frame to be limited by the minimum invariant length $l_{pl}$. Then expression (10) follows from (5), it was first got in [16, 17].

The given result concords with condition (7). Indeed, for $l_0 \to l_{pl}$: $\alpha \to 1$ and $V_{max} \to 0$. This means that a particle with the rest mass $m_0 = m_{pl}$ ("planckeon" [4]) must be in *the state of observer independent rest*. According to (9), the momentum of motion of the planckeon equals to zero, its energy equals to $E_{pl}$, and the momentum of rest equals to $p_{pl} = m_{pl}c$. Absence of the momentum of motion means that the relative motion of planckeons and substance observer does not exist [16-18, 29]. The state of observer independent rest of planckeons differs them greatly from known kinds of matter – substance and field. These objects can be referred to a specific (the third) kind of matter. Due to its properties the planckeon can be considered as a structural element of a vacuum-like medium [16-18, 29].

For substance objects (for instance, elementary particles with the rest mass $m_0$, $0 < m_0 < m_{pl}$) $0 < V_{max}^2 < c^2$. Thus, according to expressions (9), each of them is characterized not only by the maximum finite value of energy $E_{max} = m_{pl}c^2$ but also by the maximum finite value of momentum of motion

$$p_{max}^2 = \frac{m_0^2 V_{max}^2}{1 - V_{max}^2/c^2} \equiv p_{pl}^2 - p_0^2,$$

where $p_0 = m_0 c$ is the momentum of rest of the particle, $p_{pl} = m_{pl}c$ is the Planck momentum. These values are relativistically invariant (observer independent).

From expressions (9) follows the connection between the energy and the momentum:

$$E^2 - p^2 c^2 = m_0^2 c^4. \tag{11}$$

Equation (11) expresses the law of conservation of energy-momentum for all kinds of matter [16, 17]:

---

[4] Hypothetical particles with the Planck rest mass were studied by J.A. Wheeler [30], M.A. Markov [6, 7], K.P.Stanyukovich [31]. The hypothesis of the planckeon possessing the property of relativistically invariant rest was first presented in [16].

$E^2 - p^2c^2 = m_0^2 c^4$  – substance ($m_0 > 0$);
$E^2 - p^2c^2 = 0$  – photons ($m_0 = 0$);
$E^2 = m_0^2 c^4$,  – planckeons ($m_0 = m_{pl} = \text{inv}$, $p = 0$).

### 5. Concluding remarks

The suggested concept of introduction of the notion of the maximum velocity of observed motion of an object makes it possible to exclude a number of fundamental difficulties inherent in relativistic theories and, at the same time, to preserve both postulates of SR. The quantity $V_{max}$ is defined in terms of FPC and depends on physical characteristics of the object:

$$V_{max}^2 = c^2 \left(1 - \frac{l_{pl}^2}{l_0^2}\right). \qquad (12)$$

This is an observer independent quantity, although in general case it does not belong to the set of Planck quantities due to dependence on the proper parameters of the object. Thanks to the generalizing nature of ESR the notion of the maximum velocity includes invariant limits: $V_{max} = c$ when $m_0 = 0$ (photons) and $V_{max} = 0$ when $m_0 = m_{pl} = \text{inv}$ (planckeons).

In any inertial frame the observed velocity of motion of an object with the rest mass $m_0$ is limited by the value $V_{max}(m_0)$. In our view the asymptotic achievement of the maximum velocity of the particle can be to some extent analogous to limitation of the possibility to get information about the motion of the object intersecting the Schwarzschild sphere in GR.

The maximum velocity of an object arises in the variant of ESR considering all Planck quantities related to mechanics $(E_{pl}, p_{pl}, m_{pl}, l_{pl}, t_{pl}, c, \hbar, ...)$ as observer independent limits. Thereby the bounds of the field of applicability of SR are defined in an explicit form, which excludes the possibility for appearance of physically meaningless results and save from the necessity to use artificial cut-off procedures, since the theory has a set of natural relativistically invariant cut-off parameters.

Suggested variants of DSR theories ("Doubly Special Relativity") are connected with a "deformation" (generalization) of the Lotentz transformations, the dispersion relation of SR. But nonlinear transformations of coordinate and time being multichoice, it causes the problem of justification of the uniqueness of the suggested form of transformations. Deformation of the dispersion relation leads to the energy-dependent speed of light, which in its turn gives rise to a number of difficulties. Particularly, it's not clear which value of the speed of light should be used in formulae describing the motion of a certain material object. The approach to extension of SR[5] presented in this work is free from this kind of problems. The speed of light remains a universal constant, as before, and the transformations of coordinate and time keep the form of the linear Lorentz transformations. The retention of the form of the Lorentz transformations is the consequence of the use of the model of discrete-continuous space-time which assumes, as the first approximation, keeping the properties of the flat space-time of SR on all scales down to $l_{pl}$ and $t_{pl}$. The next approximation must be connected with a change of the expression for metric (2), which will be reflected directly in the mathematical form of transformations. They will obviously have to take into account the quantities $l_{pl}$ and $t_{pl}$, but it's not possible so far to predict exactly their form.

---

[5] The results presented here are not a completed variant of ESR, they just outline an approach to the development of a new theory that must be created using new mathematical concepts corresponding to the "discrete-continuous" structure of space-time. Thus the problem of studying the properties of the model of the discrete-continuous space-time and creating an adequate mathematical apparatus that would take into account existence of the invariant limiting element of the set is brought to the forefront. Development of such a mathematical tool will ensure the correct use of differentiation operations in the theory, finite values of invariant elements can cause non-linearity of arithmetical operations, similar to the specific character of velocity composition in SR [20].

From the physical point of view, the essential novelty of the theory including the concept $V_{max}$ is the description of the kinematic state of observer independent rest. Similar to the description of the invariant (observer independent) motion of photons in SR $(m_0 = 0, V_{max} = c)$, in the framework of the extended theory it's possible to introduce a concept of invariant rest of planckeons $(m_0 = m_{pl}, V_{max} = 0)$. Thereby the planckeon, as well as the photon, differs in principle from substance objects. It can be supposed that these objects (planckeons) form a medium, which is a model of the vacuum-like state of matter – relativistic ether[6]. In this case characteristic physical parameters of the given medium are the Planck quantities that manifest themselves as universal invariant limits for substance objects.

This conception allows us to substantiate SR's postulates owing to the existence of relativistic ether. Indeed, all IRF are in the same kinematic state in relation to this fundamental medium, *therefore* they are physically equivalent (the relativity principle, absence of ether "wind") and the velocity of perturbation propagation is the same in different IFR $c = \text{const}$. Existence of the relativistic medium does not contradict the basis of modern fundamental physical theories (SR, GR, Quantum Field Theory and others). At the same time non-interrelated postulates of these theories find common explanation on the ground of the idea of the relativistic medium filling up space.

## Acknowledgments

We are grateful to our good friend and colleague Prof. A.L. Simanov for his constant encouragement, helpful remarks and fruitful discussions of the issues touched on in the paper.

---

[6] Impossibility of ether detection by means of inertial motion can be connected with the peculiar equation of state of this medium [32, 33]: $P = -\varepsilon$, where $P$ is pressure, $\varepsilon$ is energy density. In this medium any IRF is locally concomitant (there is no "headwind"), in contrast to substance (for which there exists a unique locally concomitant IRF) and field (for which the concept of concomitant frame doesn't exist).